\documentstyle[12pt]{article}

\begin{document} 
\newcommand{\be}{\begin{equation}}
\newcommand{\ee}{\end{equation}}
\title{
Instantons and Non-Perturbative Dynamics  of  
$N=2$ Supersymmetric
Abelian Gauge Theories in Two Dimensions.
}
\author{
  A.A.Penin\\
  {\small {\em Institute for Nuclear Research of the
  Russian Academy of Sciences,}}\\
  {\small {\em 60th October Anniversary
  Pr., 7a, Moscow 117312, Russia.}}
  }
\date{}

\maketitle

\begin{abstract}
We study $N=2$ supersymmetric Abelian gauge model
with the  Fayet-Iliopoulos term and 
an arbitrary number of chiral matter multiplets
in two dimensions. By analyzing the instanton 
contribution we compute the non-perturbative 
corrections to  the mass spectrum of the theory 
and the quantum deformation of the classical vacua space. 
In contrast to  known  examples the non-perturbative
bosonic potential is saturated by the {\it one-instanton}
contribution and can be directly found within
the semiclassical   expansion  around the one-instanton 
saddle point.
\end{abstract}

\thispagestyle{empty}

\newpage

\section{Introduction}
The non-perturbative aspects of supersymmetric field theories
has been extensively studied  and it has become clear
that a wide class of their properties can be analyzed 
exactly. It has been also found that the  non-perturbative dynamics 
of supersymmetric theories is essentially governed by instantons.  
Thus the instanton based calculations  
result in  determination of the vacuum structure of some  
$N=1$ supersymmetric non-Abelian gauge theories in $d=4$ 
dimensions   \cite{NSVZ4,ADS,Ven}.   Similar results  have been obtained 
in $d=2$ supersymmetric sigma models \cite{NSVZ2,Nar}. 
Another example is the four-dimensional non-Abelian
gauge theories with  $N=2$ extended supersymmetry where
the exact solution for the  quantum 
moduli space is known  which sums up all terms of the 
instanton expansion \cite{SeiWit,Fin}. 
The instantons play an essential role also
in dynamics of $d=3$ supersymmetric gauge models \cite{AHW,3dim1,3dim2}. 

We consider the effects of the 
instantons in the $d=2$, $N=2$ supersymmetric
Abelian gauge theory with the  Fayet-Iliopoulos term and 
an arbitrary number of chiral matter multiplets.
The models of this type  were  considered  
in the context of 
Calabi-Yao/Landau-Ginzburg correspondence in ref.~\cite{Wit2}
where some instanton effects were announced. 
In the present paper we develop the  semiclassical 
perturbation theory (PT) around the one-instanton 
saddle point.  By analyzing the instanton 
contribution to a set of Green functions (GFs) 
we construct the effective action which describes
the effects of the instantons at large distance.
As usual the instantons violate the non-renormalization theorems
and induce the correction to the mass spectrum and 
the deformation of the classical vacua space of the model.
However, in contrast to other supersymmetric theories
the non-perturbative bosonic potential which determines
the spectrum and the vacuum structure of the model
is saturated by the {\it one-instanton}
contribution and can be directly computed  within
the PT  around the one-instanton 
solution.

\section{The model}
The $N=2$ supersymmetric
Abelian gauge model in two dimensions can  be 
obtained by dimensional reduction of the $d=4$, $N=1$
supersymmetric QED \cite{Wit2,DF}.
The model is constructed from the charged  chiral  multiplets
$\Phi^i$  ($i=1,\ldots ,N_f$) and the gauge  vector multiplet $V$.
The (anti)chiral superfields  have the following expansion in 
terms of the component fields\footnote{See Appendix A for spinor algebra notations.}
$$
\Phi^i(x,\theta)=\phi^i(y)+\sqrt{2} \overline{\theta}_c\psi^i (y)
+\overline{\theta}_c\theta F^i(y),
$$
\be
\overline{\Phi}^i(x,\theta)=\overline{\phi}^i(\overline{y})+
\sqrt{2}\overline{ \theta}{\psi}^i_c (\overline{y})
+\overline{\theta}{\theta}_c\overline{F}^i(\overline{y}),
\label{chi}
\ee
$$
y_\mu=x_\mu+i\overline{\theta}\gamma_\mu\theta .
$$
The expansion of the gauge superfield  reads
(in Wess-Zumino gauge)
\be
V = -\overline{\theta} \gamma^\mu\theta v_\mu
-\overline{\theta}{1+ \gamma_5\over 2}\theta\sigma
-\overline{\theta}{1- \gamma_5\over 2}\theta\overline{\sigma}
+i\overline{\theta}_c\theta \overline{\theta}{\lambda}_c
-i\overline{\theta}\theta_c\overline{\theta}_c\lambda
+{1\over 2}\overline{\theta}_c\theta\overline{\theta}{\theta}_cD.
\label{gau}
\ee
The (anti)chiral  superfields 
obey the constraints $\overline{{\rm D}}\Phi^i=0$,
 ${\rm D}\overline{\Phi}^i=0$
where ${\rm D}$  ${\rm \overline{D}}$  stands for 
the supercovariant derivative
$$
{\rm D} = {\partial\over\partial\overline{\theta}_c}-i
\lefteqn{\partial}{/}{\theta}_c , 
$$
\be
\overline{{\rm D}}   =- 
 {\partial\over\partial\theta_c}+i\overline{\theta}_c
\lefteqn{\partial}{/} ,
\label{Der}
\ee   
where $\lefteqn{\partial}{/}=\gamma^\mu\partial_\mu$.
In two dimensions it is possible to construct also  the twisted chiral 
multiplet  $\Sigma$   by 
acting with the supercovariant derivatives on the  vector multiplet \cite{Wit2,Gate}
$$
\Sigma= {1\over \sqrt 2}\overline{ {\rm D}}_R {\rm D}_L V(z)=
\sigma(z) + i \sqrt 2\theta_L
\overline{\lambda}_R(z) - i\sqrt 2\overline{\theta}{}_R
\lambda_L(z) -
$$
\be
-\sqrt {2}\theta_L\overline{\theta}{}_R\left(D(z)+i
{\varepsilon^{\mu\nu}v_{\mu\nu}(z)\over 2}\right),
\label{Sig}
\ee
where $v_{\mu\nu}=\partial_\mu v_\nu -\partial_\nu v_\mu $ 
and $z_\mu=x_\mu+i\overline{\theta}\gamma_\mu\gamma_5\theta$.
The  twisted chiral superfield obeys the constraints
$ {\rm D}_{L}\Sigma =\overline {{\rm D}}_{R}\Sigma =0$. The  twisted 
antichiral superfield  $\overline{\Sigma}$ is a complex conjugate 
of eq.~(\ref{Sig}) and obeys $ {\rm D}_{R}\overline{\Sigma} =
\overline {{\rm D}}_{L}\overline{\Sigma} =0$.

In the  superspace $(\theta,~\overline{\theta},~x)$ 
supersymmetry is realized  by the operators
$$
Q = {\partial\over\partial\overline{\theta}_c}+
i\lefteqn{\partial}{/}{\theta}_c , 
$$
\be
\overline{Q} =- 
 {\partial\over\partial{\theta}_c}+
i\overline{\theta}_c\lefteqn{\partial}{/} , 
\label{Q}
\ee  
with the following algebra
\be
\{Q,\overline{Q}\}=-2i\lefteqn{\partial}{/}
+Z .
\label{comm}
\ee
In eq.~(\ref{comm}) $Z$ is a complex central charge
corresponding to  the momenta in the reduced directions
of the $N=1$ algebra in four dimensions \cite{3dim2}. 

The supersymmetry transformation  of the component fields can be directly
obtained from eqs.~(\ref{chi},~\ref{gau},~\ref{Q}). However,
it is relatively complicated \cite{Wit2} because the supersymmetry 
transformations have to be accompanied by gauge transformation to preserve 
the Wess-Zumino gauge.

The Lagrangian of the model has the form \cite{Wit2,DF}
\be
L=L_{{matter}}+
L_{{ gauge}}+L_{{\zeta}},
\label{L}
\ee
where the first two terms in the right hand side  
are the kinetic energy
of the chiral fields and  the  kinetic energy
of the gauge field respectively. The third term 
is a twisted chiral  superpotential which incorporates both 
the Fayet-Iliopoulos term and  the theta angle~\cite{Wit2}.
No chiral superpotential interaction is allowed in our model 
because we choose all the 
chiral fields to be of the same gauge charge.
In terms of superfields the right hand side of eq.~(\ref{L})  reads 
\be
L_{ matter}
  =\int\!{\rm d}^2 x {\rm d}^4\theta \sum_i \overline\Phi_{i} e^{2V}
\Phi^{i}, 
\label{Lm}
\ee
\be
L_{ gauge}=-{1\over 4e^2}
\int\! {\rm d}^2x{\rm d}^4\theta \overline{\Sigma} \Sigma , 
\label{Lgauge}
\ee
\be
L_{\zeta}=
{\zeta\over 2\sqrt 2}\int\! {\rm d}^2x{\rm d}\theta_L{\rm d}
\overline{\theta}{}_R 
\left.\Sigma\right|_{\theta_R=\overline{\theta}{}_L=0}+{ h.c.},
\label{Leta}
\ee
where $e$ is a gauge coupling constant of dimension one
in mass units and $\zeta = \eta^2 - i\theta/(2\pi)$ with real  
$\eta^2$ and $\theta$.
The component expansion of the
first two terms in eq.~(\ref{L})
is  
$$
L_{matter}= \int\!{\rm d}^2x\sum_i\left(
-D^{\mu}\overline{\phi}{}^iD_{\mu}\phi^i+
i\overline{\psi}{}^i{\lefteqn{D}{\,/}}\,{\psi}^i+
\overline{F}{}^i{F}^i
-2\overline{\sigma}\sigma\overline{\phi}{}^i{\phi}^i
+D\overline{\phi}{}^i{\phi}^i-\right.
$$
\be
\left.
-\sqrt{2}{\sigma}\overline{\psi}{}^i{1+\gamma_5\over 2}{\psi}^i
-\sqrt{2}\overline{\sigma}\overline{\psi}{}^i{1-\gamma_5\over 2}{\psi}^i
+i\sqrt{2}\overline{\phi}{}^i\overline{\psi}{}^i_c\lambda
-i\sqrt{2}{\phi}^i\overline{\lambda}{\psi}{}^i_c\right) ,
\label{Lmcom}
\ee
\be
L_{gauge}= {1\over e^2}\int\!{\rm d}^2x\left(
-{1\over 4}v^{\mu\nu}v_{\mu\nu} 
-\partial^{\mu}\overline{\sigma}\partial_{\mu}\sigma+
i\overline{\lambda}{\lefteqn{\partial}{/}}\lambda+ {1\over 2}D^2 
\right) ,
\label{Lgcom}
\ee
where $D_{\mu}=\partial_{\mu}-iv_{\mu}$.
Eq.~(\ref{Leta}) in components reads 
\be
L_{\zeta}= -\int\!{\rm d}^2x\left(\eta^2 D+{\theta\over 2\pi}
{\varepsilon^{\mu\nu}v_{\mu\nu}
\over 2}\right)
\label{Letacom}
\ee
{\it i.e.} the constants
$\eta^2$ and $\theta$ parameterize  the Fayet-Iliopoulos  
and  theta angle terms respectively. 

Let us consider some properties of the model.
At the classical level it 
has left- and right-moving $R$ symmetries \cite{Wit2}
which provides us with the standard non-renormaliza\-tion
theorems \cite{W}.
There is also a chiral symmetry which rotates
$\psi^i$, $\sigma$ and  $\lambda$ fields and their
complex conjugates so that $(\psi^i_L,~\psi_R^i,~\lambda_L,~\lambda_R,~
\sigma)$  have charges $(-1,~1,-1,~1,-2)$.
Both the $R$ and chiral symmetries are anomalous at the quantum
level. Classically the chiral invariance of the Lagrangian leads
to   conservation of the axial current 
\be
J_\mu^5=\overline{\psi} \gamma_\mu\gamma_5\psi-
{1\over e^2}\overline{\lambda} \gamma_\mu\gamma_5\lambda+
{1\over e^2}(\overline{\sigma}
\partial_\mu\sigma-\sigma \partial_\mu \overline{\sigma}),
\label{acurr}
\ee
so $\partial^\mu J_\mu^5=0$. However,
non-invariance of the partition function
under the chiral transformation  results
in the ``diangle'' anomaly in the divergence of the 
axial current \cite{Sch} 
\be
\partial^\mu J_\mu^5={1\over 2\pi}N_f\varepsilon^{\mu\nu}v_{\mu\nu}. 
\label{anom}
\ee
Because of the anomaly  the  $\theta$ angle transforms under the 
chiral rotation with the parameter
$\alpha_\chi$ as follows 
\be
\theta \rightarrow \theta - 2N_f\alpha_\chi .
\label{thtr}
\ee
Thus all  values of the $\theta$ angle are equivalent
and  one can  put $\theta =0$.

The equation of motion for the auxiliary fields can
be solved with the result
$$
D=-e^2(\sum_i\overline{\phi}{}^i\phi^i -\eta^2) ,
$$
\be
F^i=0 .
\label{aux}
\ee 
After eliminating the auxiliary fields the potential 
energy takes the form 
\be
U(\phi_i,~\sigma,~v_\mu)= {e^2\over 2}\left(\sum_i\overline{\phi}{}^i\phi^i 
-\eta^2\right)^2+
\left(2\overline{\sigma}\sigma-v^\mu v_\mu\right)
\sum_i\overline{\phi}{}^i\phi^i .
\label{poten}
\ee
Thus for a non-zero value  of the parameter $\eta$  
at the classical level the Higgs effect takes place  
while the Coulomb phase is absent.
In the case $N_f=1$ no massless fields survive. 
The spectrum  consists of the 
massive vector field  and the real scalar field and there are 
two distinct vacua $\phi =\pm \eta$.
When  $N_f>1$ vanishing of $U$ requires
\be
\sum_i\overline{\phi}{}^i\phi^i =\eta^2 ,
\label{const}
\ee
{\it i.e.} the classical vacua space of the model 
is a $CP^{(N_f-1)}$ manifold with the K{\"a}hler class $\eta^2$ and 
the massless fields form a supersymmetric $CP^{(N_f-1)}$ model.
Note that for an arbitrary $N_f$ the Witten index $(-1)^F$ \cite{Witi}
is non-zero so spontaneous supersymmetry breaking does 
not happen even at the quantum level.

\section{$N_f=1$ model in detail}
\subsection{Mass spectrum and running coupling}
When $N_f=1$  and  $\eta \ne 0$ 
all the fields get masses via the Higgs mechanism. 
The mass spectrum saturates 
the Bogomol'nyi-Prasad-Sommerfield (BPS) bound \cite{WB,B,PS,WitOl}
\be 
m \ge {1\over 2}|Z| ,
\label{mass}
\ee
so both the gauge and chiral fields  are the sum of 
short multiplets \cite{3dim2}.
The central term is a  combination of
$U(1)$ charges $Y$ and $\overline{Y}$  
\be
Z=2 \sqrt{2}e\eta
\left(
\begin{array}{cc}  
Y & 0\\
0 & \overline{Y} \\
\end{array}  
\right).
\label{cench}
\ee
where the  $Y$  transformation  mixes the components of the vector
and chiral multiplets: 
$\delta\overline{\phi} \sim\sigma/e,~
\delta{\phi} \sim{\sigma}/e,~
\delta{\psi}_R \sim i\overline{\lambda}_R/e
,~\delta\overline{\psi}_L \sim i\lambda_L/e,\ldots$,
and the 
$\overline{Y}$ transformation is a Hermitian conjugate of $Y$:  
$\delta\phi \sim \overline{\sigma}/e,~
\delta\overline{\phi} \sim\overline{\sigma}/e,~
\delta\overline{\psi}_R\sim
-i\lambda_R/e,~
\delta{\psi}_L \sim -i\overline{\lambda}_L/e,\ldots$. 
Using eq.~(\ref{cench}) we find the masses of the
gauge and matter fields to be equal to   
\be
m=\sqrt{2}e\eta .
\label{classm}
\ee
Let us consider the quantum corrections
to  this  classical expression.
The PT  of the model
has a natural dimensionless expansion parameter
\be
{g^2\over  4\pi} = {1\over 4\pi\eta^2} \sim {e^2\over m^2}, 
\label{coup}
\ee
so the PT is valid 
if $\eta$ is large  enough.
In the one loop approximation the vacuum expectation 
value (VEV) $\eta$ is renormalized by the scalar tadpole
\cite{Wit2}  
\be 
\eta^2(\mu)={1\over 2\pi}\ln{\mu\over\Lambda},
\label{etarun}
\ee
where $\mu$ is an ultra-violet cutoff and $\Lambda$
parameterizes the infrared behavior of $\eta$.
Eq.~(\ref{etarun}) corresponds to the
non-zero $\beta$-function of  $g$ coupling
\be
\beta = {g^2\over 4\pi^2}\beta_0, \qquad \beta_0=1 .
\label{beta} 
\ee
Thus instead of eq.~(\ref{classm}) one gets 
the one-loop corrected expression
\be 
m^2(\mu)={1\over \pi}e^2\ln{\mu\over\Lambda}.
\label{mrun}
\ee
In fact the superrenormalizability of the model and the standard  
non-renormalization theorems  tell  us that
eq.~(\ref{mrun}) is the exact quantum analog of eq.~(\ref{classm})
at least within the PT framework \cite{Wit2}.  

It is important that  
the currents associated with the left- and right-moving
$R$ symmetries are also anomalous because 
they couple to  the chiral charged fermions.
Since the $R$ symmetry is anomalous the non-renormalization 
theorems do  not work in the instanton sector. As a consequence
eq.~(\ref{mrun}) can be modified by instantons.

\subsection{ Instanton solution and instanton measure}
The described model is known  to support  the instanton  solutions --
the Abrikosov-Nielsen-Olesen vortices \cite{DF,Ab}\footnote{The analytical
continuation from the Minkowski to Euclidean metric is implied 
(see Appendix~A)}.
The one-instanton solution is spherically symmetric
and in the polar coordinates ($\alpha_r,~r$) can be parameterized as follows 
\be 
\begin{array}{c}  
v^I_\mu =\varepsilon_{\mu\nu}\partial_\nu \Psi_v(r),\qquad 
\partial_\nu \Psi_v(0) = 0, \\ 
\phi_I =e^{i\alpha_r}\eta(1 -\Psi_\phi(r)),\qquad  \Psi_\phi(0)=1.\\
\end{array}  
\label{inst}
\ee
where the functions $\Psi_\phi(r)$ and $\Psi_v(r)+\ln({Cer})$
exponentially decay at large $r$ (the constant  $C$ is  a
characteristic of the solution). Meanwhile, there is no 
analytical expression of  $\Psi_\phi(r)$ and $\Psi_v(r)$ 
in terms of elementary functions.

The instanton configuration   
satisfies the first-order Bogomol'nyi equations \cite{B}
\be
\begin{array}{l}  
(D-v_{01})_I=0,\\
(iD_0+D_1)\phi_I=0.\\
\end{array}  
\label{bogom}
\ee
As a consequence   the instanton of the winding number $N$ 
\be
N={1\over 2\pi}\int \!{\rm d}^2xv_{01}
\label{topch}
\ee
has the  action $2\pi N\eta^2$ and 
is neutrally stable with respect to  
$N$ instantons of the winding number $1$ \cite{B,ENS}.

In the instanton background the Dirac operator
acting on the $\psi$ and $\lambda$ fields
has zero modes \cite{TH}. It is convenient to arrange the 
$\psi_L$ and $\overline{\lambda}{}_R$ fields into one two-component spinor
\be
\chi =
\left(
\begin{array}{c}  
\psi_L\\
\overline{\lambda}{}_R/e \\
\end{array}\right) , 
\ee
so the (Euclidean) Dirac operator on these components takes the form
\be 
i\lefteqn{D}{\,/}\,=
\left(
\begin{array}{cc}  
-D_0+iD_1 & -i\sqrt{2}\phi \\
i\sqrt{2}\overline{\phi}&  -\partial_0-i\partial_1\\ 
\end{array}  
\right) .
\label{dir}
\ee
Using  the Bogomol'nyi equations
it is possible  to show   that
zero modes of positive  chirality do not exist
if  the external field configuration
has a positive winding number \cite{DF}. 
Then the chiral anomaly tells us that in the one-instanton 
background there is one complex zero mode of negative
chirality \cite{NS}. 
This mode is generated by the  (complex) supersymmetry
transformation of the instanton field \cite{Zu}
\be
\chi^0 =
\begin{array}{cc}  
\left(
\begin{array}{c}  
\psi_L^0\\
\overline{\lambda}{}_R^0/e \\
\end{array}\right)=
\left(
\begin{array}{c}  
\sqrt{2}(iD_0-D_1)\phi_I \\
(D+v_{01})_I/e \\
\end{array}\right).
\end{array}
\label{hfzmod}
\ee
The rest of the supersymmetry generators
does not affect the instanton field
because of eq.~(\ref{bogom}).

Now it is straightforward to compute
the one-instanton contribution to the
partition function -- the instanton measure.
The contribution of the two  real bosonic  zero modes
associated with the translation invariance of the 
instanton solution to the measure is \cite{GST} 
\be
S_I\int\!{\rm d}^2x_0 , 
\label{bmes}
\ee
where the collective coordinate $x_0$
corresponds to the position of the instanton and  
the factor $S_I$ normalizes the zero modes.
The part of the measure corresponding to the 
one complex fermionic zero mode reads \cite{TH}
\be
J^{-1}\int\!{\rm d}\zeta_0{\rm d}\overline{\zeta}_0 ,
\label{fmes}
\ee
where $\zeta_0$ is the Berezin integration variable
and the factor 
\be
J=4S_I
\label{J}
\ee
normalizes the zero mode~(\ref{hfzmod}) to $1$.
The contribution from the bosonic and fermionic quadratic functional
integration around the instanton solution due to the 
non-zero modes cancel each other because of supersymmetry \cite{DD}.
As usual,  the renormalization scale
$\mu$  appears in  the measure 
as a contribution of the Pauli-Villars regulator fields
to the fermionic and bosonic determinants due to the zero modes \cite{TH}.
The power of $\mu$ is exactly the first coefficient
of the $\beta$-function~(\ref{beta}) 
$\beta_0=1$ which  measures the difference between the
numbers of bosonic   and fermionic zero modes.   

Multiplied by  the exponent of the one-instanton action
eqs.~(\ref{bmes},~\ref{fmes}) give the final expression for the measure 
\be
{\Lambda\over 4}\int\!{\rm d}^2x_0{\rm d}
\zeta_0{\rm d}\overline{\zeta}_0,
\label{Z}
\ee
where
\be
\Lambda = \mu e^{-2\pi\eta^2(\mu)}.
\label{Lam}
\ee
Note that the  PT corrections
to this result exist due to the Yukawa coupling
\be
-\sqrt{2}\overline{\sigma}\overline{\psi}_R\psi_L .
\label{Yucoup}
\ee

\subsection{Green functions and  effective Lagrangian}
Let us  consider the following GFs 
\be
G_{\chi\overline{\chi}}(x-y)=\langle 0|{\chi (x)}
{\overline{\chi}(y)}|0\rangle,
\label{Gcc}
\ee
\be
G_{\sigma}(x)=\langle 0|{\sigma} (x)|0\rangle .
\label{Gs}
\ee
These GFs vanish within the PT and have
relevant chiral transformation properties to get the one-instanton 
contribution.
 
Using the instanton measure~(\ref{Z})
one finds the GF~(\ref{Gcc}) to be saturated by
the fermionic zero mode~(\ref{hfzmod})
\be
G_{\chi\overline{\chi}}(x-y)
= {\Lambda\over 4}\int\!{\rm d}^2x_0\,\chi^0(x-x_0)
\overline{\chi}{}^0(y-x_0) . 
\label{Gcci}
\ee 
From eqs.~(\ref{bogom},~\ref{hfzmod},~\ref{Gcci}) one obtains 
the non-zero condensate of the  $\psi_L$ and 
$\overline{\lambda}{}_R$ fermionic components 
\be
{1\over e^2}\langle 0|\lambda_L\overline{\lambda}{}_R
|0\rangle = 
\langle 0|\overline{\psi}{}_R{\psi_L}|0\rangle
= \pi\eta^2\Lambda .
\label{lcon}
\ee 
Eq.~(\ref{Gcci}) implies also that in the one-instanton 
approximation  a non-local 
diagonal mass matrix  for the   $\psi_L$ and $\overline{\lambda}{}_R$
components appears
\cite{AHW}
\be
\delta \hat m(x-y)=\int \!{\rm d}^2x\, \lefteqn{D}{\,/}\,^0_x
G_{\chi\overline{\chi}}(x-y)\overleftarrow{\lefteqn{D}{\,/}}{}^0_y ,
\label{mnloc}
\ee
where ${\lefteqn{D}{\,/}}\,^0$ is the free Dirac operator
\be 
i{\lefteqn{D}{\,/}}\,^0=
\left(
\begin{array}{cc}  
-\partial_0+i\partial_1 & -i\sqrt{2}\eta \\
i\sqrt{2}\eta & -\partial_0 -i\partial_1 \\ 
\end{array}  
\right) .
\label{dirfr}
\ee
Near the mass shell eq.~(\ref{mnloc}) can be reduced up to the higher derivatives
to the effective local  mass matrix
\be
\delta \hat m=\int \!{\rm d}^2x\, \hat m(x) .
\label{mmloc}
\ee
Using eqs.~(\ref{bogom},~\ref{topch},~\ref{hfzmod}) in the local
limit one finds the effective real masses 
for the $\lambda$ and  $\psi$ fields
\be
\delta m_{\lambda\overline{\lambda}} = 
\delta m_{\overline{\psi}\psi} = 8\eta^2\pi^2\Lambda \equiv \delta m .
\label{mloc}
\ee
The fact that eqs.~(\ref{lcon},~\ref{mloc}) are equivalent for the 
$\psi$ and $\lambda$ fields is a consequence of the non-anomalous
$Y$ ($\overline Y$)  symmetry. 
Because of the standard non-renormalization theorem \cite{NSVZ2}
the PT corrections to the two-point GF 
of the gauge fermion  and therefore to the parameter
$\delta m_{\lambda\overline{\lambda}}$  are absent.
On the other hand, the  PT corrections to 
the two-point GF of the matter fermion  are possible
due to the Yukawa coupling~(\ref{Yucoup}). However, taking into account
the $Y$  ($\overline Y$) symmetry one concludes that these corrections are
also absent unless the symmetry is broken.  

The calculation of the instanton contribution 
to the GF~(\ref{Gs}) is a little bit more complicated. 
It vanishes at the tree level because of the fermionic zero mode 
so one has to take into account the 
loop corrections  to the partition function. 
Then one gets\footnote{See Appendix B.} 
\be
G_{\sigma}(x)=
4\sqrt{2}\eta^2\pi^2\Lambda
\equiv\sigma_0 .
\label{scon}
\ee
Thus the instantons lead to the condensation of the lowest
component of the vector superfield. A non-zero VEV
of the $\sigma$ field violates the $Y$  ($\overline Y$)
symmetry and induces the  corrections to the  
two-point GF of the matter fermion field. 
We will consider these corrections below.

The contribution of the  anti-instanton solution with
a negative winding number to the  partition function 
is a Hermitian conjugate of the 
instanton contribution. For example, the anti-instantons
saturate the condensates and real mass terms for  the 
${\lambda}{}_R$ and  $\overline{\psi}{}_L$ fermionic components.

Now we are able to write down the effective Lagrangian to 
describe the effects induced by the (anti-)instantons 
at large distance.
This Lagrangian has to reproduce 
eqs.~(\ref{mloc},~\ref{scon}) and  respect 
all non-anomalous symmetries of the fundamental Lagrangian~(\ref{L}).
These  constraints are satisfied with the Lagrangian  of the following
unique  form    
\be
\delta L=
\int\! {\rm d}^2x{\rm d}\theta^4 
\left(\overline{\Phi} e^{2(V-\tilde V)}\Phi-
\overline{\Phi} e^{2V}\Phi\right)  
+{\delta m\over 4e^2}\left(\int\! {\rm d}^2x{\rm d}\theta_L
{\rm d}\overline{\theta}{}_R \left.
\Sigma^2
\right|_{\theta_R=\overline{\theta}{}_L=0}
+{ h.c.}\right),
\label{Leffsf}
\ee
where $\tilde V$ is an auxiliary external vector superfield with
the non-zero scalar component 
$\tilde\sigma =\overline{\tilde{\sigma}}=\sigma_0$.
In components eq.~(\ref{Leffsf}) reads (in the Minkowski metric) 
$$
\delta L=\int\!{\rm d}^2x\left(
-2(\overline{\sigma}-
\overline{\tilde{\sigma}})(\sigma-\tilde\sigma)
\overline{\phi}\phi+2\overline{\sigma}\sigma\overline{\phi}\phi 
+{\sqrt{2}}\tilde\sigma
\overline{\psi}_L\psi_R+{\sqrt{2}}\overline{\tilde{\sigma}}
\overline{\psi}_R\psi_L\right)-
$$
\be
-{\delta m\over e^2}\left(\int\!{\rm d}^2x
\left({\sqrt{2}\over 2}{\sigma}\left(D-iv_{01}\right)
+\overline{\lambda}_R\lambda_L\right) +{ h.c.}\right).
\label{Leff}
\ee
The potential energy now is of the following form
\be
U(\phi,~\sigma,~v_\mu)= {e^2\over 2}
\left(\overline{\phi}\phi - 
{\sqrt{2}\over 2}{\delta m \over e^2}(\sigma +\overline{\sigma})
-\eta^2\right)^2+\left(2(\overline{\sigma}-\overline{\tilde{\sigma}})
(\sigma-\tilde\sigma)-v^\mu v_\mu\right)
\overline{\phi}\phi .
\label{poten1}
\ee
Eq.~(\ref{poten1}) determines the mass spectrum  and 
the vacuum configuration in the one-instanton approximation.
The VEV of the $\sigma$
field is given by eq.~(\ref{scon}). The VEV of the $\phi$
field can be written in the gauge invariant form
\be
\langle 0|\overline{\phi}\phi|0\rangle = \eta^2 + 
\sqrt{2}{\delta m\sigma_0\over e^2}.
\label{phicon}
\ee
Then the instanton contribution leads to 
the mass splitting: a half of the states is of  
the mass $m+\delta m/2$ while another states 
are of the mass  $m-\delta m/2$ (where we keep
only the leading order term in $\Lambda$).  
Note that the central charge~(\ref{cench})
is modified by  eq.~(\ref{Leffsf}).

These results need some comments.\\
i) As it was pointed the expressions~(\ref{mloc},~\ref{scon})
for the parameters  $\sigma_0$ and $\delta m$ are exact within the 
PT around the one-instanton solution.  
Thus  the perturbative corrections
to the partition function~(\ref{Z}) are reduced to 
the condensation of the $\sigma$ field. 
Meanwhile, the effective 
Lagrangian~(\ref{Leff}) can be modified by  multi-instanton
contributions.\\
ii) Taking into account the real mass term in eq.~(\ref{Leff})  
one can reproduce within the effective theory the value 
of  the gauge fermion condensate~(\ref{lcon}).\\
iii) Formally, the first part of eq.~(\ref{Leff}) contains 
the real mass term for the $\psi$ field with mass being equal to
$\sqrt{2}\sigma_0$. This term is a local approximation of 
the non-local mass term implied by the contribution of the 
fermionic zero mode to the two-point GF~(\ref{Gcc}) 
(see eqs.~(\ref{mloc},~\ref{scon})). However, when the 
correct vacuum configuration    
$\sigma = \sigma_0$ is used this mass term disappears. 
By the same reason  the condensate of  the 
matter fermion vanishes. In this way 
the PT corrections to the two-point GF of the matter fermion field
(eqs.~(\ref{lcon},~\ref{mloc})) are taken into account 
within the effective theory framework.\\
iv) Eq.(\ref{Leff}) includes a  bosonic part and 
in particular contains the non-diagonal mass term which mixes the 
bosonic components of the vector and chiral superfields. 
The effective coupling in front of the bosonic term is of order 
$\Lambda\sim e^{S_I}$ and therefore is saturated by the  one-instanton 
contribution. This term of the effective Lagrangian  
can be directly found by analyzing the one-instanton
contribution to the corresponding bosonic correlator.  
In contrast to the considered model 
in all known examples  
the non-perturbative bosonic potential
is suppressed by an extra power of $\Lambda$. In this case it
can  be only   reconstructed by supersymmetry transformation 
from the fermionic part of the effective action while the dynamical
calculations are not available.\\ 
v) Though eq.~(\ref{phicon}) 
is obtained by minimizing  the  effective
potential  in the one-instanton approximation there is no instanton
contribution to this quantity
because of the chiral selection rule. The non-perturbative 
variation of the  $VEV$
of the $\phi$ field  is of order $\Lambda^2$ 
so it corresponds to the contribution of the 
instanton--anti-instanton pair rather then the contribution of the
single (anti-)instanton.\\ 
vi) Naively, the  effective
potential  in the one-instanton approximation has an additional
vacuum state with 
$$  
\phi =0 ,
$$
\be
\sigma = - {e^2\eta^2\over \sqrt{2}\delta m}.
\label{flvac}
\ee  
This, however, contradicts to the constraint on the number
of the bosonic vacuum states imposed by the 
Witten index analysis. The reason of this inconsistency
is that eq.~(\ref{Leff}) is valid only for the region 
$\sigma\sim 0$ and $\phi\sim\eta$ where the PT
around the one-instanton solution is applicable.
For the large $\sigma$ small $\phi$ region
the  anomalous twisted superpotential is also known
\cite{Wit2,DDDS} 
$$
{1\over 4\sqrt{2}\pi}\int\! {\rm d}^2x{\rm d}\theta_L
{\rm d}\overline{\theta}{}_R \left.
\Sigma\ln{\left(\Sigma /\mu\right)}
\right|_{\theta_R=\overline{\theta}{}_L=0}
+{ h.c.}=
$$
\be
=-{1\over 4\pi}\int\! {\rm d}^2x
\left((\ln{\left(\sigma /\mu\right)}+1)(D-iv_{01})
+\sqrt{2}{\overline{\lambda}_R\lambda_L\over\sigma}\right)-{ h.c.}
\label{lsep}
\ee
In particular
for $\eta^2>0$ the effective potential
 has no additional vacuum states so
eq.~(\ref{flvac}) is a spurious vacuum.    
However  this  is an interesting and non-trivial  problem to find
the exact solution which interpolates between eq.~(\ref{Leff})
and eq.~(\ref{lsep}).

\section{Generalization to $N_f>1$}
If $N_f>1$ the presence of the massless  particles 
leads to the infrared divergence 
of the PT. To regularize the divergence
we add to the Lagrangian a  mass term of the form
$$
\int\! {\rm d}^2x{\rm d}\theta^4 \sum^{N_f}_{i=2}
\overline{\Phi}{}^i \left(e^{-
(m_r\overline{\theta}{}_L\theta_R+
m_r^*\overline{\theta}{}_R\theta_L)/\sqrt{2}}-1\right)\Phi^i=
$$
\be
=- \sum^{N_f}_{i=2}\int\!{\rm d}^2x\left(m_r^*m_r\overline{\phi}{}^i\phi^i
+m_r^*\overline{\psi}{}^i_R\psi_L^i+
m_r\overline{\psi}{}^i_L\psi_R^i\right),
\label{reg}
\ee
where $m_r$ is a (complex)  parameter. 
Thus $\phi^1$ becomes the only field with a non-zero VEV. 
This term breaks  $SU(N_f)$ flavor symmetry
of the original Lagrangian to $SU(N_f-1)$. 
The  chiral symmetry is not broken by eq.~(\ref{reg})
if we assume $m_r$ to be of  the chiral charge $-2$. 
Then  eq.(\ref{etarun}) transforms to 
\be 
\eta^2(\mu)={1\over 2\pi}\ln{\left(\mu^{N_f}\over
\Lambda |m_{r}|^{N_f-1}\right)},
\label{etarun1}
\ee
so  $\beta_0=N_f$.
The instanton solution for $N_f>1$ is identical to one of the
$N_f=1$ model: the gauge field and the scalar field
$\phi^1$ are given by eq.~(\ref{inst}) while $\phi^i$ 
$(i=2,\ldots,N_f)$ are equal to zero\footnote{If some of the
chiral superfields have opposite gauge charges the theory
has a non-trivial moduli space and the structure of 
the instanton solution becomes much more complicated \cite{PRTT}.}.
However, the instanton measure is  modified. At $m_r=0$ 
the Dirac operator in the one-instanton background in addition to
eq.~(\ref{hfzmod}) has $N_f-1$ extra massless fermionic zero modes
\be
\psi_L^i = e^{\Psi_v}, \qquad \psi_R^i=0, \qquad i=2,\ldots,N_f .
\label{lfzmod}
\ee
These modes can not be obtained by a supersymmetry transformation
of the instanton field. On the other hand they are related 
by the supersymmetry transformation to the additional bosonic
zero modes
\be
\phi^i = e^{\Psi_v}, \qquad i=2,\ldots,N_f .
\label{lbzmod}
\ee
Though these modes are non-normalizable they do contribute to the 
instanton measure because the numbers of the bosonic and fermionic modes
are equal \cite{BKN}. At small $m_r$ the contribution of the zero
modes~(\ref{lfzmod}) to the fermionic determinant is ${m_r}^{(N_f-1)}$
 up to 
the higher order corrections in $m_r$. Taking into account the 
contribution of the regulator fields we find the total contribution of the
fermionic zero modes~(\ref{lfzmod}) to the instanton measure to be    
\be
\left({m_r\over \mu}\right)^{(N_f-1)} .
\label{fcont}
\ee
Similarly the bosonic zero modes~(\ref{lbzmod}) give the factor 
\be
\left({\mu^2\over m_r^*m_r}\right)^{(N_f-1)} .
\label{bcont}
\ee
The contribution of the translation and supersymmetry zero modes 
is  the same as in the $N_f=1$ case. So the
total power of $\mu$ is equal to  $\beta_0$.
Putting all the factors together and 
taking into account eq.~(\ref{etarun1}) we found that the instanton measure
for an arbitrary $N_f$ is the same as in the $N_f=1$ model and is 
given by   eq.~(\ref{Z}) up to the factor
$e^{i(N_f-1)\alpha_m}$ where $\alpha_m$ is a phase of $m_r$. 
This phase, however, can be absorbed by the redefinition of the
$\theta$ angle  which enters the instanton measure as the factor
$e^{i\theta}$
so that under the chiral group it transforms
according to eq.~(\ref{thtr}) (another contribution to
eq.~(\ref{thtr}) comes from the supersymmetric fermionic zero mode). 
The only difference between the models with 
different $N_f$ is in the expression of the running coupling $\eta^2$
so we can extend the results of the previous section to an arbitrary $N_f$.
In particular we find that the instanton contribution  results in 
the non-perturbative variation~(\ref{phicon})
of the K{\"a}hler 
class of the $CP(N_f)$ model describing the light sector of the model.

The limit  of a vanishing $m_r$ needs a comment. 
In this limit one expects the phase transition 
since Goldstone bosons do not exist in two dimensions \cite{Col}.
Moreover, the $CP(N_f)$ model of the massless fields 
at $m_r=0$ supports  the ``light'' instanton solutions which involve only 
the light degrees of freedom \cite{DDL}.
We, however,  do not consider these effects and  are rather interested 
in the calculation of  the Wilsonian effective action for
the massless fields by integrating out the ``heavy'' instantons of
the fundamental theory \cite{Wit2}.  
As we see the instanton measure  does not depend on $m_r$ explicitly. 
Then  the chiral transformation properties of the
instanton measure (of the fermionic determinant) is not changed in the
limit $m_r=0$ since  ${N_f-1}$  zero modes eq.~(\ref{lfzmod})
appear instead of the factor ${m_r}^{(N_f-1)}$.
Thus the parameter $m_r$
in eq.~(\ref{etarun1}) can be considered as some infrared cutoff 
$\sim\Lambda $ of the 
Wilsonian effective action without specifying a regularization 
procedure.

\section{Conclusion}
We have analyzed the non-perturbative 
properties of the $N=2$ supersymmetric Abelian gauge 
theory in two dimensions.
We have developed the  systematic semiclassical expansion around
the  one-instanton  saddle point and presented the set of 
the bosonic and fermionic GFs saturated by the
one-instanton contribution. The fermionic GFs 
are determined  by the zero modes of the Dirac operator
in the instanton field background while one has to take 
into account the  loop  corrections to the
partition function to compute the   
bosonic field correlators.  
We also constructed the effective action  
which describes the instanton 
induced effects and explicitly reproduces these GFs 
at large distance.
It includes   the non-perturbative 
contribution to the K{\"a}hler potential and the non-perturbative
twisted superpotential.
In contrast to 
all known models  the effective coupling constant 
parameterizing the  bosonic part of this action 
turns out to be of the first  order in 
$\Lambda\sim e^{S_{inst}}$ and is saturated by the 
one-instanton contribution. 
This potential
determines the  corrections to the   
BPS saturated  mass spectrum 
of the model and  to the classical vacuum configuration
of the scalar fields.  In particular it implies 
the non-vanishing VEV of the  
scalar component of the vector superfield.
Then the instanton contribution 
results in the non-perturbative variation of the K{\"a}hler 
class of the $CP^{(N_f-1)}$ manifold of  
the classical vacua space.

\vspace{5mm}

\noindent
{\large \bf Acknowledgments}\\[1mm]
I  am grateful to V.A.Rubakov, P.G.Tinyakov and S.V.Troitsky  for numerous 
valuable discussions.
This work  is supported in part by Volkswagen Foundation under contract
I/73611 and  by Russian Basic Research Foundation
under contract  97-02-17065.

\section{Appendix A}
We use the following conventions in two dimensions.
The metric tensor is $g_{\mu\nu}={\rm diag}(1,~-1)$ and  
the antisymmetric tensor $\varepsilon_{\mu\nu}$ is defined so that
$\varepsilon_{01}=-\varepsilon^{01}=1$. 
The gamma matrices are in the following representation     
\be
\begin{array}{lcr}  
\gamma_0 = \left(
\begin{array}{cc}  
0 & 1\\
1 & 0\\
\end{array}  
\right),&
\gamma_1 = \left(
\begin{array}{cc}  
0 & -1\\
1 & 0 \\
\end{array}  
\right),&
\gamma_5 =-\gamma_0\gamma_1 =\left(
\begin{array}{cc}  
-1 & 0\\
0 & 1 \\ 
\end{array}  
\right).\\
\end{array}  
\label{gamma}
\ee
so  a spinor $\theta$ and its Dirac conjugate 
$\bar\theta =\theta^\dagger\gamma_0$ are of the form   
\be 
\begin{array}{cc}  
\theta = \left(
\begin{array}{c}  
\theta_L \\
\theta_R \\
\end{array}  
\right),&
\overline{\theta} = \left(
\begin{array}{cc}  
\overline{\theta}{}_R &
\overline{\theta}{}_L \\
\end{array}  
\right).\\
\end{array}  
\label{spin}
\ee
where $\overline{\theta}{}_R=\theta_R^*$,  $\overline{\theta}{}_L=\theta_L^*$.
The charge conjugated spinors are defined as follows
$$
\theta_c=\gamma_c\overline{\theta}{}^T
$$
\be
\overline{\theta}{}_c= \theta^T\gamma_c
\label{chcon}
\ee
where $\gamma_c=\gamma_1$ is  the  charge conjugation  matrix
$\gamma_c\gamma_\mu\gamma_c=-\gamma_\mu^T$.

To study  the instanton solution we make the analytic continuation 
from the Minkowski to
Euclidean metric  $x_0\rightarrow -ix_0$.
The Euclidean Dirac matrices are
$$
\gamma_0^E=-i\gamma_0, \qquad \gamma_1^E=\gamma_1,
$$
\be 
\gamma_5^E=-i\gamma_0^E\gamma_1^E=\gamma_5,\qquad \gamma_c^E=\gamma_c.
\label{gammae}
\ee
The representation~(\ref{spin}) is still valid 
but in Euclidean space 
one has to substitute the Dirac conjugation by the Hermitian one
$\bar\theta =\theta^\dagger$ so 
$\overline{\theta}{}_R=\theta_L^*$,
$\overline{\theta}{}_L=\theta_R^*$.

\section{Appendix B}
GF~(\ref{Gs}) can be computed within the standard
PT around the one-instanton solution \cite{Ven}
(note that the fermionic zero  mode is cancelled by the Yukawa 
coupling~(\ref{Yucoup})). 
In this way, however,
one needs to know the 
propagator of the $\sigma$
field in the one-instanton background \cite{Ven}.
Fortunately, it is possible to bypass this dynamical problem by the
following trick. Using the functional integral representation
of the partition function 
it is straitforward to prove a ``low energy theorem'' 
(see, for example, \cite{Wit3}) 
\be
{d\over d{\zeta}}G_{\sigma}(x)=-\int \!
{\rm d}^2y\langle 0|{\sigma}(x)
\left(D(y)+v_{01}(y)\right)|0\rangle .
\label{lowth}
\ee 
On the other hand, the bosonic  correlator
\be
\langle 0|{\sigma}(x)
\left(D(y)+v_{01}(y)\right)|0\rangle
\label{GsD}
\ee
is related to the function
\be
\langle 0|{\overline{\lambda}{}_R(y)}{\lambda_L(x)}|0\rangle
\label{Gll}
\ee
by the supersymmetric Ward-Takahashi
identity \cite{NSVZ4,Ven}. Indeed, using the transformation law 
\be
\left[\overline{Q}{}_L,{\sigma}\right]=-i\sqrt{2}\lambda_L, \qquad
\{\overline{Q}{}_L,\overline{\lambda}_R\}=-i(D+v_{01})
\label{com}
\ee
and the fact  that the vacuum state is
annihilated by the supersymmetry generators
because  supersymmetry is not spontaneously broken  
one finds
\be
\langle 0|{\sigma}(x)
\left(D(y)+v_{01}(y)\right)|0\rangle=-\sqrt{2}
\langle 0|{\lambda_L(x)}
{\overline{\lambda}{}_R(y)}|0\rangle .
\label{GsDGl}
\ee
Substituting  the bosonic correlator 
by the fermionic one in eq.~(\ref{lowth})
according to this equation and using 
eqs.~(\ref{bogom},~\ref{topch},~\ref{hfzmod},~\ref{Gcci}) one finds
\be
{d\over d{\zeta}}G_{\sigma}(x)= 
4\sqrt{2}\pi^2\Lambda .
\label{der}
\ee 
Integrating this equation  for $\theta =0$
one obtains 
\be
G_{\sigma}(x)= 
4\sqrt{2}\eta^2\pi^2\Lambda + C'.
\label{scon2}
\ee 
where $C'$ is an integration constant. To fix this constant
one needs some boundary condition.
In the limit  $\eta^2\rightarrow0$ the theory has no Higgs phase
and the instantons smoothly transform to the pure
gauge. So one can suppose  all instanton saturated quantities
to vanish in this limit. Then $C'=0$ and
one gets  eq.~(\ref{scon}). 
The value $C'=0$ is also consistent
with the Witten index analysis. 
Indeed, if $C'\ne 0$ the real mass
term~(\ref{mloc}) of the $\psi$ field can not be reabsorbed by
the shift of the $\sigma$ field to its VEV (see the comment after  
eq.~(\ref{Leff})). Then the term 
\be
-(2{C'}^2\overline{\phi}\phi
+\sqrt{2}C'\overline{\psi}_R\psi_L+
\sqrt{2}C'\overline{\psi}_L\psi_R)
\ee  
appears in  effective Lagrangian~(\ref{Leff}) and  leads to 
spontaneous supersymmetry breaking.

Because of the non-renormalization property
of the two-point GF of the gauge fermion field 
the result~(\ref{scon}) is exact within the PT
around the one-instanton solution.

\end{document}